\documentclass[]{spie} 

\usepackage[english]{babel} 
\usepackage{amsmath,amsfonts,stmaryrd,amssymb} 
\usepackage{graphicx}
\usepackage{url}
\usepackage{hyperref}
\usepackage[table,xcdraw]{xcolor}
\graphicspath{ {images/} }
\usepackage{subcaption}
\usepackage{authblk}
\usepackage{array}
\usepackage{multirow}
\title{Exploring Fine Subpixel Spatial Resolution of Hybrid CMOS Detectors}
\begin{document}
\author[1]{Evan Bray}
\author[1]{Abraham Falcone}
\author[1]{Samuel V. Hull}
\author[1]{David N. Burrows}
\affil[1]{Pennsylvania State University}
\maketitle 
\begin{abstract}
When an X-ray is incident onto the silicon absorber array of a detector, it liberates a large number of electrons, which tend to diffuse outward into what is referred to as the charge cloud. This number can vary from tens to thousands across the soft X-ray bandpass (0.1 - 10 keV). The charge cloud can then be picked up by several pixels, and forms a specific pattern based on the exact incident location of the X-ray. We present experimental results on subpixel resolution for a custom H2RG with 36$\mu$m pixels, presented in Bray 2018\cite{Bray2018}, and compare the data to simulated images . We then apply the model simulation to a prototype small pixel hybrid CMOS detector (HCD) that would be suitable for the \textit{Lynx} X-ray surveyor. We also discuss the ability of a small pixel detector to obtain subpixel resolution.
\end{abstract}
\section{Introduction}
\begin{figure}[h]
\centering
\includegraphics[width = 0.4\textwidth]{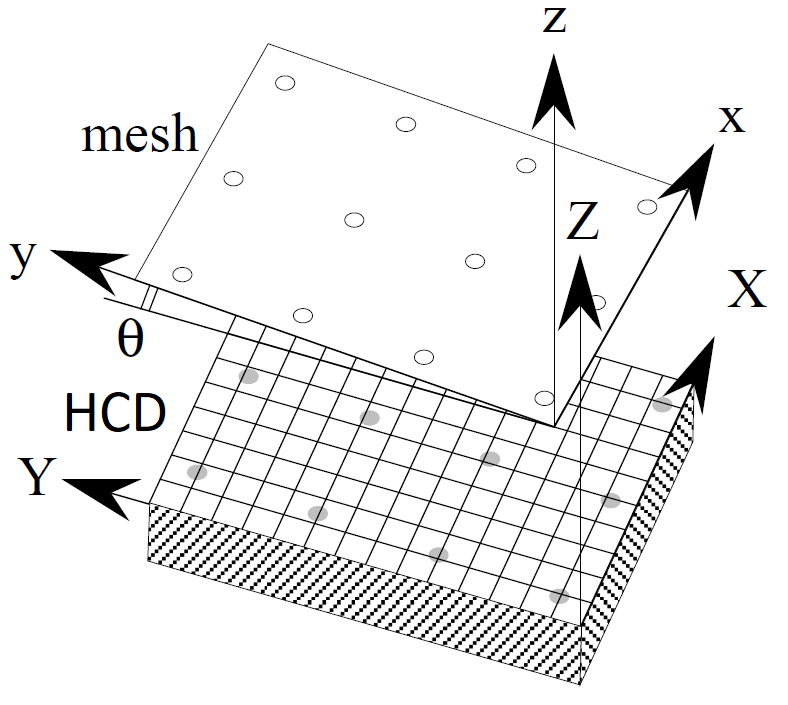}
\caption{An illustration of the mesh experiment, adapted from Tsunemi \textit{et al.} (1998)\cite{Tsunemi1998}. The mesh hole separation is a factor of four times larger than the H2RG pixel pitch. Copyright 1998 The Japan Society of Applied Physics.}
\label{fig:multi-pitch}
\end{figure}

By performing a ``mesh experiment," detector features can be measured on a subpixel scale by placing a thin metal mesh with evenly spaced holes directly over the detector, as indicated in Figure \ref{fig:multi-pitch}. The shape of the electron charge cloud can then be experimentally determined from the resulting events\cite{Tsunemi1997SPIE, Tsunemi1997, Hiraga1998}. This procedure was utilized for the Advanced CCD Imaging Spectrometer (ACIS) prior to launch on board the \textit{Chandra X-ray Observatory}\cite{Pivovaroff2000}. It has been demonstrated through on-orbit observations that utilizing a subpixel event repositioning algorithm can improve the size of the point-spread function (PSF) for on-axis sources by $\sim$50$\%$ \cite{Li2004}. This capability has played a key role in the imaging of objects like galactic nuclei\cite{Wang2011} and SN1987A\cite{Park2002}, as well as more efficient source identification in observations like Chandra Deep Field-North that contain hundreds of individual point sources\cite{Xue2016}. Utilizing this algorithm has since become a default parameter in the Chandra Interactive Analysis of Observations (CIAO) processing tool.

Taking advantage of subpixel spatial resolution will continue to be an important factor for future high resolution X-ray space telescopes, such as the \textit{Lynx} X-ray Surveyor. By combining large collecting area with high angular resolution, \textit{Lynx} will observe significantly deeper than the \textit{Chandra} while continuing to distinguish between closely-separated sources that are identified by telescopes like \textit{JWST}. At its sensitivity limit, \textit{JWST} is expected to detect $\sim$2$\times$10$^{6}$ galaxies/deg$^{2}$\cite{Windhorst2006}, and accurately matching these with faint X-ray sources out to z$\sim$10 will require that \textit{Lynx} maintains a 0.5$^{\prime\prime}$ angular resolution in order to avoid source confusion. By utilizing knowledge of the charge cloud size, it is possible to restrict the X-ray landing location to a subpixel region; effectively increasing the spatial resolution of the entire imaging array.

\section{H2RG Results} 
\subsection{Detector Description}
The charge cloud size for Al K$\alpha$ X-rays was experimentally determined for a Teledyne HyViSi H2RG to be well-represented by a 2D Gaussian with a FWHM of 13.4$\mu$m\cite{Bray2018}. The detector is composed of 1024$\times$1024 pixels with a 36$\mu$m pitch and 100$\mu$m depletion depth, and is bonded to 1 of every 4 pixels in the standard H2RG readout integrated circuit. It exhibits an energy resolution of 2.7$\%$ at 5.9 keV, and a read noise of 6.8 e$^{-}$ $\pm$ 0.1 e$^{-}$ when paired with a Teledyne cryogenic SIDECAR$^\textnormal{{TM}}$. These are the operating parameters that will be used in modeling subpixel resolution for this detector.

\subsection{Simulating Charge Cloud Size} \label{sec:Simulating Charge Cloud Size H2RG}
In the absence of experimental results from a mesh experiment, there exist other methods for determining charge cloud size. One method is to use a series of equations that describe the motion of an ensemble of electrons in a depleted semiconductor. A thorough explanation of this method is given in Yousef 2011\cite{Yousef2011}. The second method is to utilize the event type branching ratios that are determined through uniform illumination by an X-ray source. By simulating event type branching ratios, one can fit the charge cloud to a size that best matches the observed results. We find that the results of both techniques agree well with our experimentally determined value of the charge cloud size in the H2RG (a 2D Gaussian with $\sigma$=5.7$\mu$m), and apply these methods to the small pixel HCD discussed in Section \ref{sec:Small Pixel HCDs}.

\subsection{Determining Subpixel Resolution} \label{sec:Determining Subpixel Resolution H2RG}
\begin{figure}
  \centering
  \includegraphics[height = 6cm,angle=0]{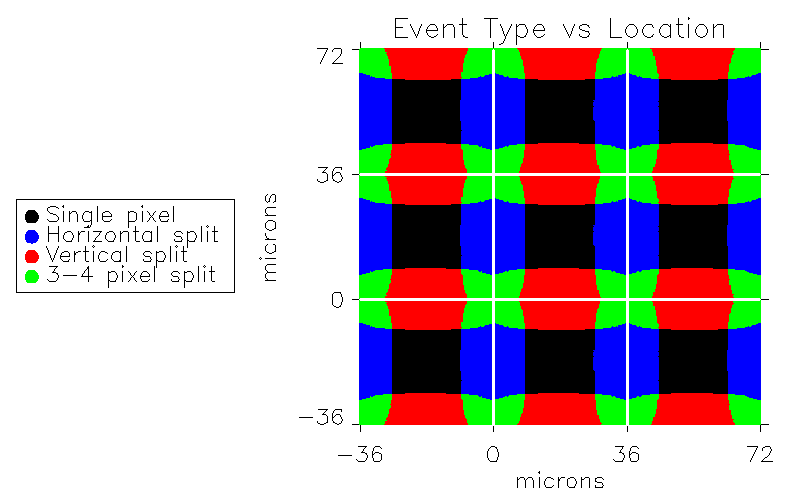}
  \caption{Simulated plot of event type vs X-ray landing location. A 3$\times$3 grid of H2RG pixels is shown to help visualize the size of each area.}
\label{fig:event_locations_H2RG}
\end{figure}

We evaluate three characteristic types of events in the following analysis; a single pixel event, horizontal split, and 3-pixel corner split event. These events are produced in the H2RG when an X-ray lands in the center, right, and bottom-right portions of the pixel, respectively. An image showing the simulated areas where various event types are produced is shown in Figure \ref{fig:event_locations_H2RG}, along with a quantitative summary of event type branching ratios in Table \ref{table:branching ratios}. A summary of the resulting subpixel resolution for the H2RG is shown in Table \ref{table:subpixel resolution summary}.

In the case of single pixel events, the only statement that can be made about the incident location of the X-ray is that it was not close enough to a pixel border to share a significant amount of charge. The results of the mesh experiment and branching ratio simulation both conclude that single pixel events are created within a 20$\times$20$\mu$m region in the center of the pixel, with a uniform probability distribution. This corresponds to a 68$\%$ confidence region having a half-width of 7.1$\mu$m in both the X and Y-direction.

\begin{figure}
\begin{subfigure}{.5\textwidth}
  \centering
  \includegraphics[height = 6cm,angle=0]{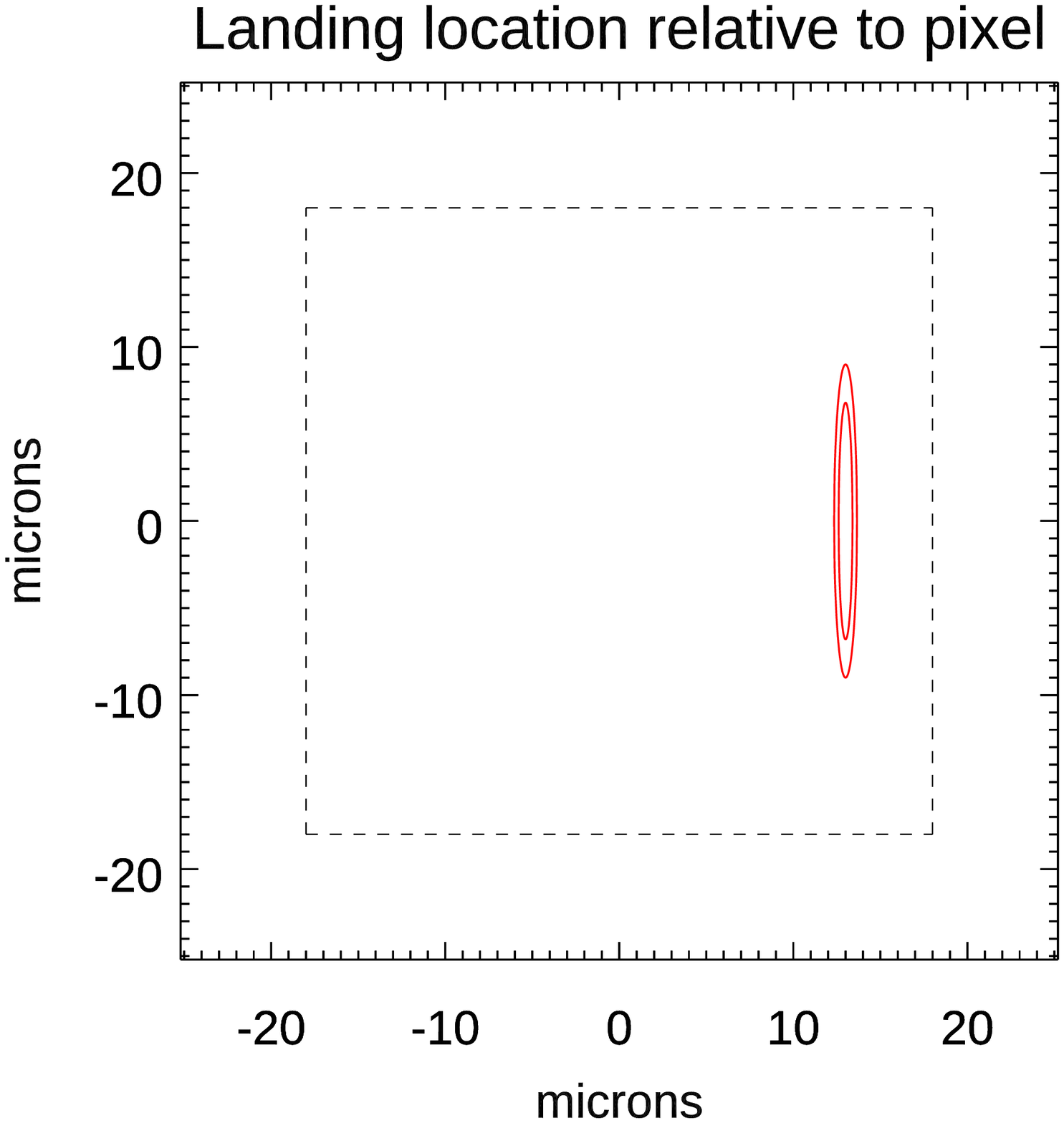}
\end{subfigure}
\begin{subfigure}{.49\textwidth}
  \centering
  \includegraphics[height = 6cm,angle=0]{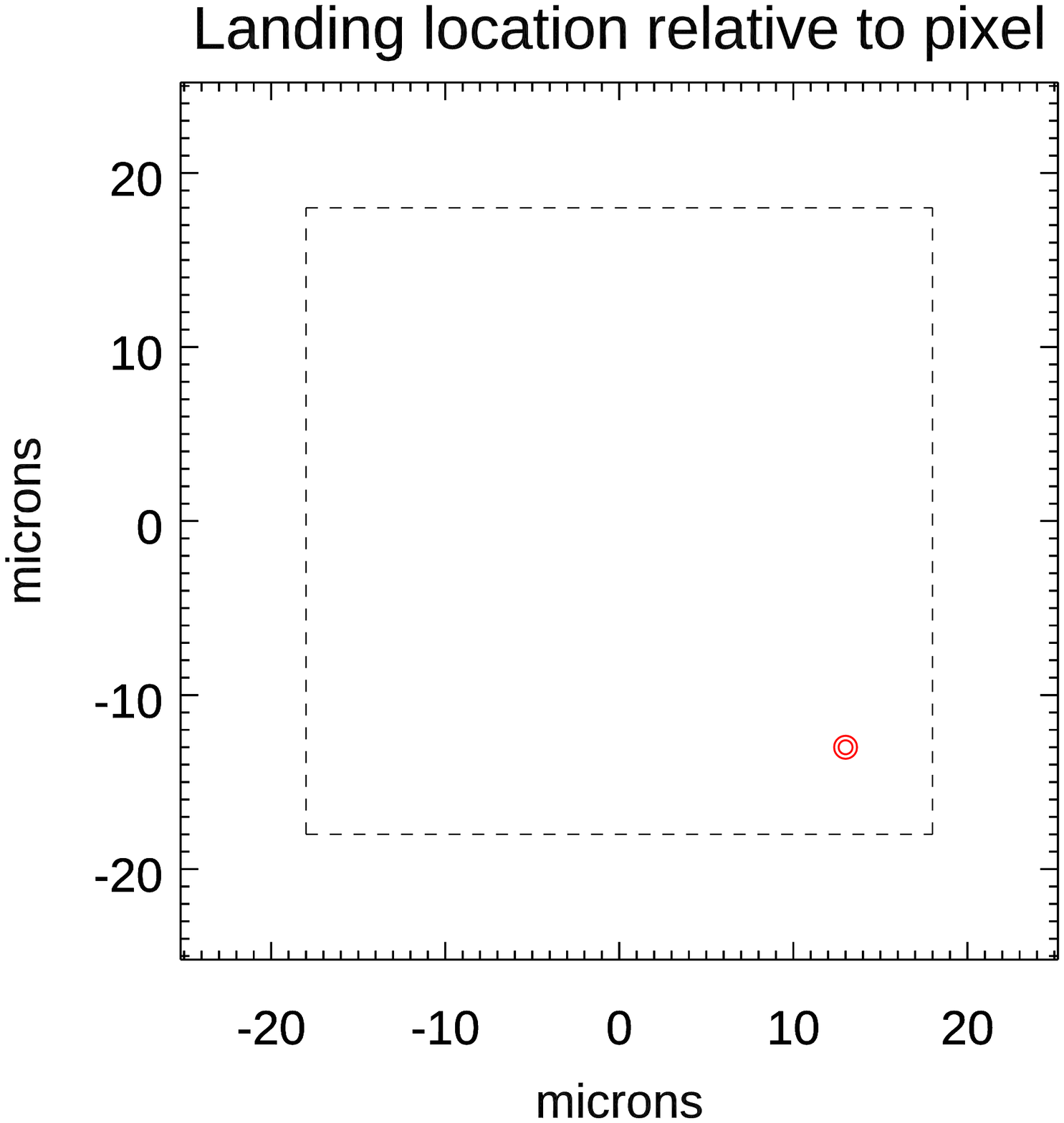}
\end{subfigure}
  \caption{The regions within the H2RG pixel to which the example horizontal split (left) and 3-pixel corner split (right) events can be localized. The 68$\%$ and 90$\%$ confidence contours are shown inside of a 36$\mu$m pixel indicated by the dashed line.}
\label{fig:refined_location_H2RG}
\end{figure}

For events that share a significant amount of charge between two or more pixels, we perform a Markov Chain Monte Carlo (MCMC) to fit the charge cloud centroid to a location that best reproduces the individual observed event. For the purposes of this analysis, we assume that each event contains an amount of signal equal to what is produced by a single Al K$\alpha$ X-ray. The resulting best fit locations for a particular horizontal split and 3-pixel corner split event are shown in Figure \ref{fig:refined_location_H2RG}. The horizontal split event shares 20$\%$ of its charge with an adjacent pixel on the right, and the 68$\%$ confidence region has a half-width of 0.4$\mu$m and 7.1$\mu$m in X and Y-directions, respectively. The 3-pixel corner split event shares 15$\%$ of its charge with adjacent pixels to the right and below, and the 68$\%$ confidence region has a half-width of 0.4$\mu$m in both the X and Y-directions.

\begin{table}
\centering
\begin{tabular}{l|ccc|}
\cline{2-4}
 & \multicolumn{3}{c|}{Experimental Branching Ratio} \\ \cline{2-4} 
 & \multicolumn{1}{c|}{\textbf{H2RG}} & \multicolumn{1}{c|}{\textbf{\begin{tabular}[c]{@{}c@{}}Small pixel HCD\\ (V$_\textnormal{{sub}}$=80V)\end{tabular}}} & \textbf{\begin{tabular}[c]{@{}c@{}}Small pixel HCD\\ (V$_\textnormal{{sub}}$=80V)\end{tabular}} \\ \hline
\multicolumn{1}{|l|}{\textbf{Pixel Pitch:}} & \multicolumn{1}{c|}{36$\mu$m} & \multicolumn{1}{c|}{12.5$\mu$m} & 12.5$\mu$m \\ \hline
\multicolumn{1}{|l|}{\textbf{Event Type}} & \multicolumn{1}{l}{\cellcolor[HTML]{C0C0C0}} & \multicolumn{1}{l}{\cellcolor[HTML]{C0C0C0}} & \multicolumn{1}{l|}{\cellcolor[HTML]{C0C0C0}} \\ \hline
\multicolumn{1}{|l|}{Single pixel} & \multicolumn{1}{c|}{36$\%$} & \multicolumn{1}{c|}{12$\%$} & 1$\%$ \\
\multicolumn{1}{|l|}{2 pixel split} & \multicolumn{1}{c|}{50$\%$} & \multicolumn{1}{c|}{45$\%$} & 2$\%$ \\
\multicolumn{1}{|l|}{3-4 pixel split} & \multicolumn{1}{c|}{14$\%$} & \multicolumn{1}{c|}{42$\%$} & 32$\%$ \\
\multicolumn{1}{|l|}{5 pixel split} & \multicolumn{1}{c|}{0$\%$} & \multicolumn{1}{c|}{1$\%$} & 21$\%$ \\
\multicolumn{1}{|l|}{6 pixel split} & \multicolumn{1}{c|}{0$\%$} & \multicolumn{1}{c|}{0$\%$} & 35$\%$ \\
\multicolumn{1}{|l|}{7 pixel split} & \multicolumn{1}{c|}{0$\%$} & \multicolumn{1}{c|}{0$\%$} & 9$\%$ \\ \hline
\end{tabular}
\caption{Event type branching ratios for the H2RG and small pixel HCD at two different levels of substrate voltage.}
\label{table:branching ratios}
\end{table}

\begin{table}[]
\centering
\begin{tabular}{l|cc|cc|cc|}
\cline{2-7}
 & \multicolumn{6}{c|}{68$\%$ Confidence region half-width ($\mu$m)} \\ \cline{2-7} 
 & \multicolumn{2}{c|}{\textbf{H2RG}} & \multicolumn{2}{c|}{\textbf{\begin{tabular}[c]{@{}c@{}}Small pixel HCD\\  (V$_\textnormal{{sub}}$=80V)\end{tabular}}} & \multicolumn{2}{c|}{\textbf{\begin{tabular}[c]{@{}c@{}}Small pixel HCD\\  (V$_\textnormal{{sub}}$=15V)\end{tabular}}} \\ \hline
\multicolumn{1}{|c|}{\textbf{Pixel Pitch:}} & \multicolumn{2}{c|}{36$\mu$m} & \multicolumn{2}{c|}{12.5$\mu$m} & \multicolumn{2}{c|}{12.5$\mu$m} \\ \hline
\multicolumn{1}{|c|}{\textbf{\begin{tabular}[c]{@{}c@{}}X-ray landing location\\ within pixel\end{tabular}}} & x & y & x & y & x & y \\ \hline
\multicolumn{1}{|l|}{Center} & 7.1 & 7.1 & 1.2 & 1.2 & 0.6 & 0.6 \\
\multicolumn{1}{|l|}{Right} & 0.4 & 7.1 & 0.2 & 1.2 & 0.4 & 0.4 \\
\multicolumn{1}{|l|}{Bottom} & 7.1 & 0.4 & 1.2 & 0.2 & 0.4 & 0.4 \\
\multicolumn{1}{|l|}{Bottom-right} & 0.4 & 0.4 & 0.2 & 0.2 & 0.4 & 0.4 \\ \hline
\end{tabular}
\caption{A quantitative summary of the subpixel resolutions that can be achieved with the H2RG and small pixel HCD. X-ray landing location can be constrained very well in directions that exhibit charge sharing with neighboring pixels.}
\label{table:subpixel resolution summary}
\end{table}

\section{Small Pixel HCDs} \label{sec:Small Pixel HCDs}
\subsection{Detector Description}
A new type of small pixel HyViSi HCD was designed through a collaboration between Teledyne Imaging Systems and the Pennsylvania State University. The purpose of these detectors is to satisfy the desire for small pixel size and high frame rates for mission concepts like the \textit{Lynx} X-ray Surveyor, while simultaneously implementing features like in-pixel correlated double sampling (CDS), and CTIA amplifiers that eliminate interpixel capacitance. The detectors are composed of 128$\times$128 pixels with a 12.5$\mu$m pitch and 200$\mu$m depletion depth. They have exhibited an energy resolution of 2.67$\%$ at 5.9 keV, and a read noise of 5.5$\pm$0.1 electrons\cite{Hull2018}. It is worth noting that both of these characteristics are the best ever measured for an X-ray HCD. 

\subsection{Simulating Charge Cloud Size}
 Because a mesh experiment was not performed for these detectors, we rely on the simulation methods described in Section \ref{sec:Simulating Charge Cloud Size H2RG} to determine the charge cloud size. The small pixel HCDs are typically operated at a substrate voltage of either 15V or 80V, so we explore both cases in the following analysis. We determine charge cloud size to be well-represented by a 2D Gaussian with a $\sigma$ of 8.2$\mu$m and 3.0$\mu$m for 15V and 80V modes, respectively, and find that both methods of simulation are in good agreement with one another.

\subsection{Simulating Subpixel Resolution}
In the following analysis, we compare the cases of the detector being operated at a substrate voltage of 15V and 80V by analyzing events that are produced by X-rays landing in the center, right, and bottom-right portions of the pixel. Although the charge cloud is spread between more pixels when a lower substrate voltage is applied, which leads to an improved subpixel centroid position, the energy resolution is worsened by the inclusion of more pixels, each with significant read noise. By utilizing the same methods described in Section \ref{sec:Determining Subpixel Resolution H2RG}, we demonstrate that significant charge spreading isn't necessary in order to achieve considerable improvement in spatial resolution for 12.5$\mu$m pixels. The experimentally determined event type branching ratios are listed in Table \ref{table:branching ratios}, while a summary of the calculated subpixel resolution is shown in Table \ref{table:subpixel resolution summary}.

\begin{figure}
\begin{subfigure}{.42\textwidth}
  \centering
  \includegraphics[height = 6.0cm,angle=0]{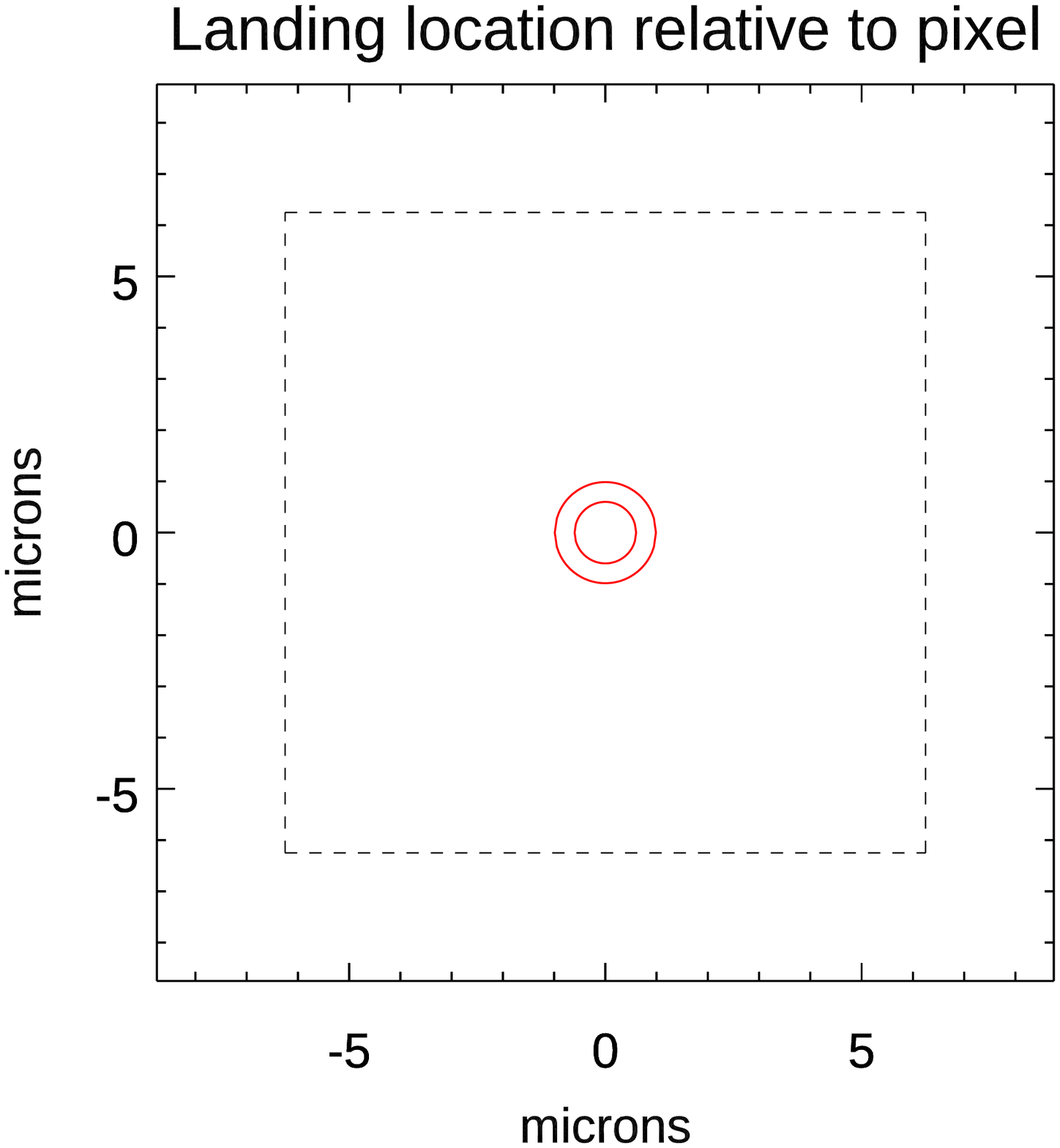}
\end{subfigure}
\begin{subfigure}{.57\textwidth}
  \centering
  \includegraphics[height = 6.0cm,angle=0]{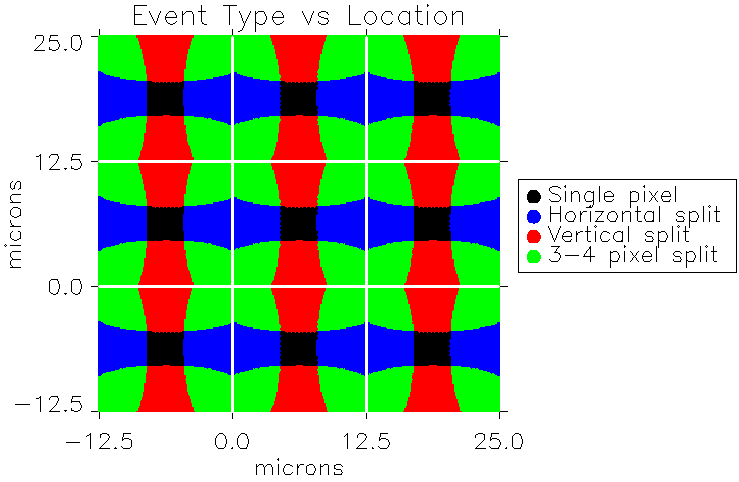}
\end{subfigure}
  \caption{(Left) The region within the pixel to which the centrally-incident X-ray can be localized for the small pixel HCD operated at 15V. The 68$\%$ and 90$\%$ confidence contours are shown inside of a 12.5$\mu$m pixel indicated by the dashed line. (Right) Simulated plot of event type vs X-ray landing location for the small pixel HCD operated at 80V. A 3$\times$3 grid of 12.5$\mu$m pixels is shown to help visualize the size of each area.}
\label{fig:SP center events}
\end{figure}

For X-rays that land near the center of the pixel, the bias voltage determines whether or not the charge cloud spreads over multiple pixels. At 15V, X-rays incident on the center of the pixel produce a 5-pixel cross-shaped event. Utilizing the charge cloud centroid method for these events returns a 68$\%$ confidence region with a half-width of 0.6$\mu$m in both the X and Y directions. At 80V, there exists a 3.4$\times$3.4$\mu$m region in the center of the pixel that produces single pixel events, which corresponds to the best fit location having a  68$\%$ confidence region with a half-width of 1.2$\mu$m in both directions. The best fit location for centrally-incident X-rays at 15V and simulated areas for event type branching ratios at 80V are shown in Figure \ref{fig:SP center events}.

\begin{figure}
\begin{subfigure}{.50\textwidth}
  \centering
  \includegraphics[height = 6.0cm,angle=0]{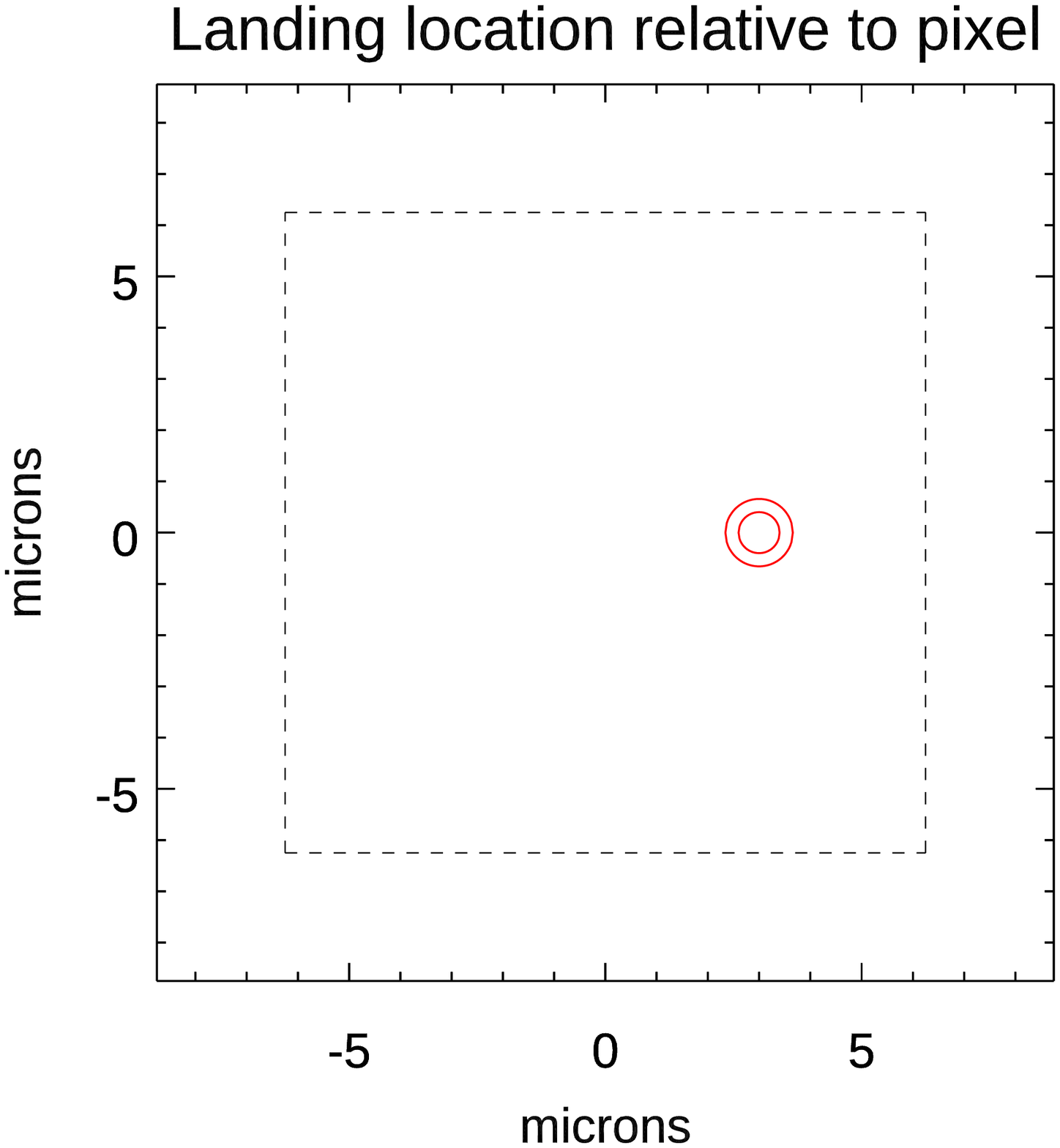}
\end{subfigure}
\begin{subfigure}{.49\textwidth}
  \centering
  \includegraphics[height = 6.0cm,angle=0]{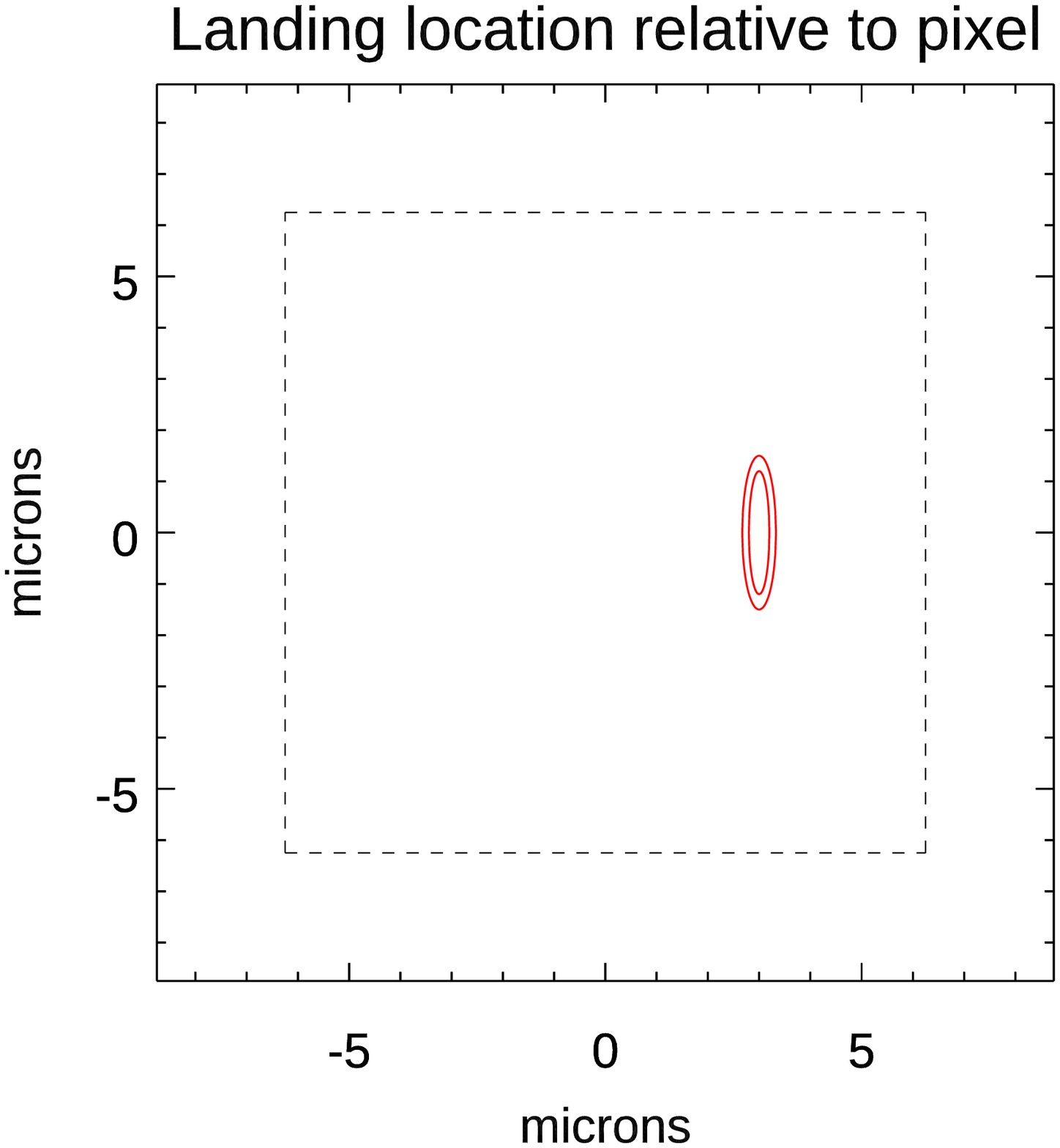}
\end{subfigure}
  \caption{The region within the small pixel HCD pixel to which X-ray landing position can be localized, for both 15V (Left) and 80V (Right). In both cases, the incident X-ray was simulated to have landed 3$\mu$m to the right of pixel center. The 68$\%$ and 90$\%$ confidence contours are shown inside of a 12.5$\mu$m pixel indicated by the dashed line.}
\label{fig:SP right events}
\end{figure}

\begin{figure}
\begin{subfigure}{.50\textwidth}
  \centering
  \includegraphics[height = 6.0cm,angle=0]{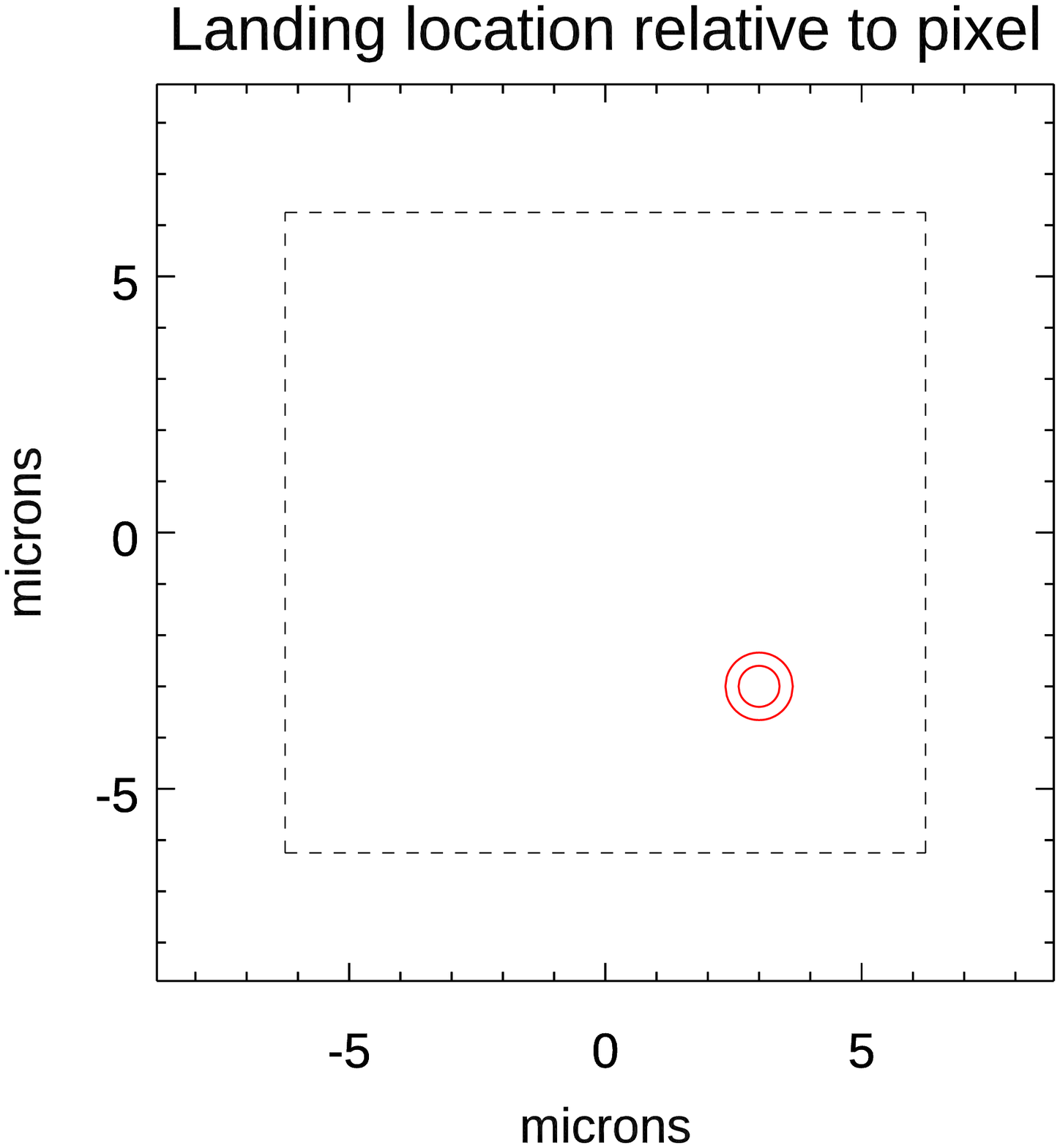}
\end{subfigure}
\begin{subfigure}{.49\textwidth}
  \centering
  \includegraphics[height = 6.0cm,angle=0]{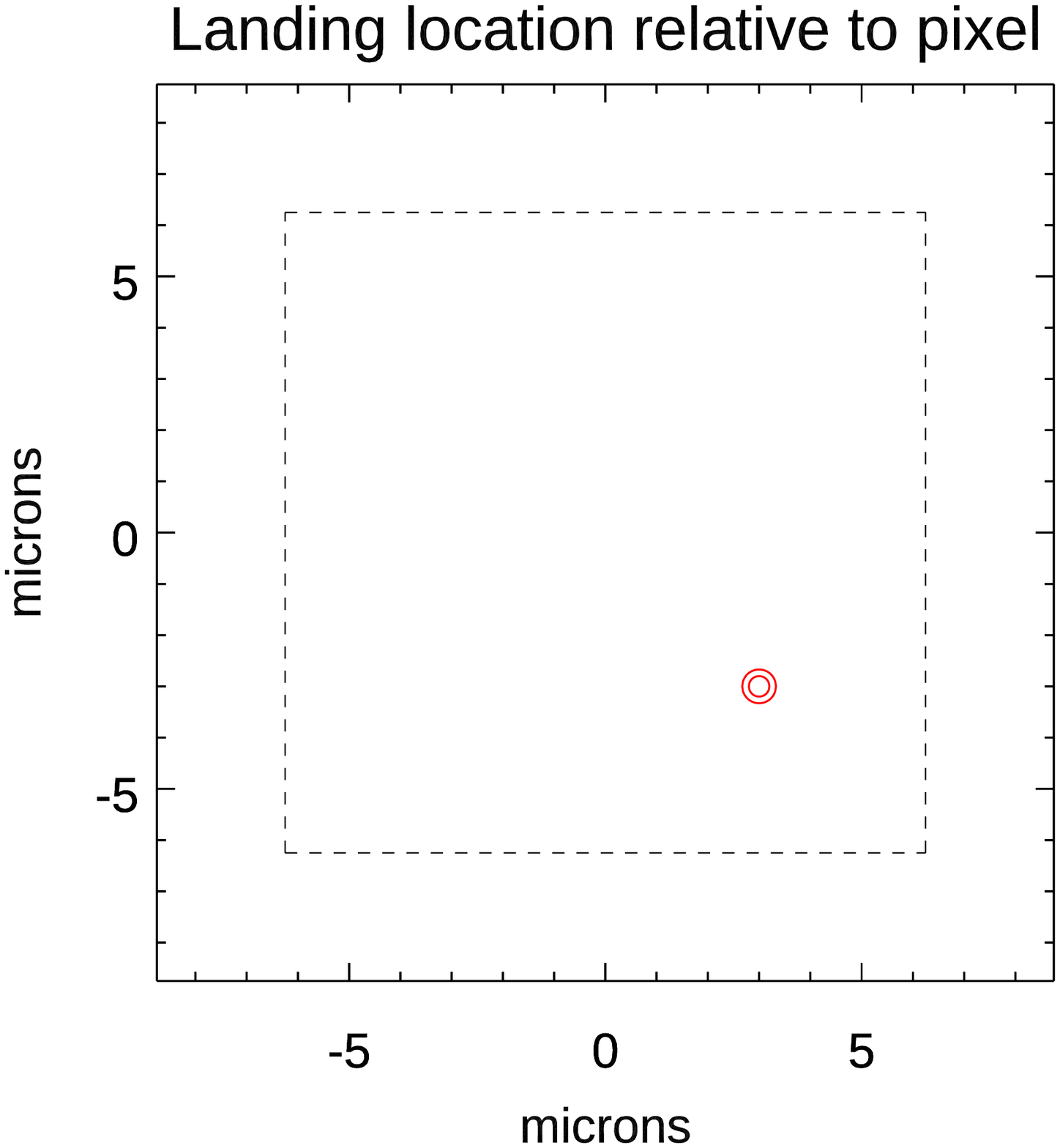}
\end{subfigure}
  \caption{The region within the small pixel HCD pixel to which X-ray landing position can be localized, for both 15V (Left) and 80V (Right). In both cases, the incident X-ray was simulated to have landed 3$\mu$m to the right, and 3$\mu$m below pixel center. 68$\%$ and 90$\%$ confidence contours are shown inside of a 12.5$\mu$m pixel indicated by the dashed line.}
\label{fig:SP bottom-right events}
\end{figure}

We also analyze events that occur from an X-ray landing in the right (bottom-right) portions of the pixel, as shown in Figure \ref{fig:SP right events} (Figure \ref{fig:SP bottom-right events}). For operation at 80V, this results in the charge cloud being spread over two (three) pixels, while at 15V the charge cloud is spread over seven (six) pixels. At 80V, this results in a 68$\%$ confidence region with half-widths of 0.2$\mu$m (0.2$\mu$m) and 1.2$\mu$m (0.2$\mu$m) in the X and Y-directions, respectively. At 15V, this results in a 68$\%$ confidence region with half-widths of 0.4$\mu$m (0.4$\mu$m) in both directions.

\section{Conclusion} \label{sec:conclusion}
By utilizing knowledge of the electron cloud size produced by each individual photon, X-ray astronomy has the ability to infer more information about the incident location of each X-ray. In order to achieve the science goals set forth by the NASA Astrophysics Roadmap, future X-ray space telescopes like the \textit{Lynx} X-ray Surveyor, will need to maintain the high level of angular resolution set forth by \textit{Chandra}. By analyzing some characteristic Al K$\alpha$ events for two types of detectors, we find that significant improvements in spatial resolution can be achieved by employing a charge cloud centroid technique. For a custom H2RG with 36$\mu$m pixels, we demonstrate a spatial resolution of 0.4$\mu$m (68$\%$ confidence) is obtained for events that share charge between three or more pixels (14$\%$ of events), and we show that the more typical 1 and 2 pixel events (86$\%$ of events) exhibit between 0.4$\mu$m and 7.1$\mu$m spatial resolution (68$\%$ confidence), which also indicates a major benefit from subpixel centroiding. We also explored the potential for increased spatial resolution in a new type of small pixel HCD with 12.5$\mu$m pixels, and found that a higher substrate voltage produces minor improvements in subpixel spatial resolution for 3-4 pixel split events by decreasing the average size of each charge cloud. When operated at 80V, we determine a spatial resolution of 0.2-1.2$\mu$m (68$\%$ confidence) for events that spread charge over two or more pixels (88$\%$ of events). When operated at 15V, we determine a spatial resolution of 0.4-0.6$\mu$m (68$\%$ confidence) for all event types. These results indicate that there exists a ``sweet spot" in charge cloud size vs pixel size that produces the best all-around results for a majority of event types. This knowledge will be useful in the design and development of future small pixel HCDs that are being investigated for use on future X-ray space telescopes.  

\section*{Acknowledgments}
This work was supported by NASA grants NNX14AH68G, NNX16AO90H, NNX16AE27G, 80NSSC18K0147, NNX13AE57G, and NNX16AE27G.

\clearpage
\bibliography{Exploring_Subpixel_Resolution}{} 
\bibliographystyle{unsrt}
\end{document}